\documentclass[11pt]{article}
\RequirePackage[a4paper,margin=2.5cm]{geometry}
\usepackage{graphicx}
\usepackage{amsmath,bm,hepunits}
\usepackage{hyperref}
\usepackage{cite}

\usepackage[utf8]{inputenc}

\usepackage{color}

\begin{document}

\begin{titlepage}

\begin{center}

\begin{flushright}
LU-TP 20-08, DESY 20-024, Nikhef 2020-004
\\
TIF-UNIMI-2020-7, Cavendish-HEP-19/03, IPPP/19/7
\\
TTK-19-03, TUM-HEP-1184/19
\end{flushright}

\vspace{3ex}

\textbf{ \LARGE Top-quark pair production at complete-NLO accuracy with NNLO+NNLL$'$ corrections in QCD }

\vspace{5ex}

\textsc{ Micha\l\ Czakon$^a$, Andrea Ferroglia$^{b,c}$, Alexander Mitov$^d$, Davide Pagani$^{e,f}$, Andrew S. Papanastasiou$^d$, Benjamin D. Pecjak$^g$, Darren J. Scott$^{h,i}$, Ioannis Tsinikos$^j$, Xing Wang$^{k}$, Li Lin Yang$^{l}$, Marco Zaro$^{m}$ }

\vspace{2ex}

\textsl{
$^a$Institut f\"{u}r Theoretische Teilchenphysik und Kosmologie, RWTH Aachen University, D-52056 Aachen, Germany
\\[0.3cm]
$^b$New York City College of Technology, Brooklyn, NY 11201, USA
\\[0.3cm]
$^c$The Graduate School and University Center, The City University of New York, New York, NY 10016  USA
\\[0.3cm]
$^d$Cavendish Laboratory, University of Cambridge, CB3 0HE Cambridge, UK
\\[0.3cm]
$^e$Technische Universit\"at M\"unchen, James-Franck-Str. 1, D-85748 Garching, Germany
\\[0.3cm]
$^f$Theory Group, Deutsches Elektronen-Synchrotron (DESY), D-22607 Hamburg, Germany
\\[0.3cm]
$^g$Institute for Particle Physics Phenomenology, University of Durham, DH1 3LE Durham, UK
\\[0.3cm]
$^h$Institute for Theoretical Physics, University of Amsterdam, Science Park 904, 1098 XH Amsterdam, The Netherlands
\\[0.3cm]
$^i$Nikhef, Theory Group, Science Park 105, 1098 XG, Amsterdam, The Netherlands
\\[0.3cm]
$^j$Theoretical Particle Physics, Department of Astronomy and Theoretical Physics, Lund University, S\"olvegatan 14A, SE-223 62 Lund, Sweden
\\[0.3cm]
$^k$PRISMA+ Cluster of Excellence, Johannes Gutenberg University, D-55128 Mainz, Germany
\\[0.3cm]
$^l$Zhejiang Institute of Modern Physics, Department of Physics, Zhejiang University, Hangzhou 310027, China
\\[0.3cm]
$^m$INFN, Sezione di Milano, Via Celoria 16, 20133 Milano, Italy
}
\end{center}

\vspace{2ex}

\begin{abstract}
We describe predictions for top-quark pair differential distributions at hadron colliders, which combine state-of-the-art NNLO QCD calculations and NLO electroweak corrections together with double resummation at NNLL$'$ accuracy of threshold logarithms and small-mass logarithms. 
This is the first time that such a combination has appeared in the literature.
Numerical results are presented for the invariant-mass distribution, the transverse-momentum distribution as well as rapidity distributions.
\end{abstract}

\end{titlepage}

\section{Introduction}

Top-quark pair production is one of the most important processes at the Large Hadron Collider (LHC). It allows one to precisely study the properties of the top quark, which are related to many important questions in particle physics, such as the hierarchy problem, the stability of the electroweak vacuum, as well as the origin of fermion masses. Top-quark pair production is also a major background in searches for many rare processes in the Standard Model (SM) and in new physics models beyond the SM.

Currently, the most precise fixed-order calculation in Quantum ChromoDynamics (QCD) for top-quark pair production reaches the next-to-next-to-leading order (NNLO) \cite{Baernreuther:2012ws, Czakon:2012zr, Czakon:2012pz, Czakon:2013goa, Czakon:2014xsa, Czakon:2015owf, Czakon:2016dgf, Catani:2019iny, Catani:2019hip}. Despite the high precision of the NNLO result, the complicated kinematics of $t\bar{t}$ production makes it necessary to consider even higher order corrections. In particular, this is due to the fact that the large collider energy at the LHC enables the study of boosted top-quark pairs, where the energies of the top quarks are much larger than their rest mass $m_t$. In \cite{Czakon:2016dgf}, it has been found that the NNLO QCD differential cross sections in the boosted regime are rather sensitive to the choice of factorization and renormalization scales. This scale dependence can be dramatically reduced by resumming certain towers of large logarithms to all orders in the strong coupling $\alpha_s$ \cite{Czakon:2018nun}. These include not only the threshold logarithms which arise when the partonic center-of-mass energy approaches the $t\bar{t}$ invariant mass $M_{t\bar{t}}$, but also the small-mass logarithms of the form $\ln^n(m_t^2/M_{t\bar{t}}^2)$ which develop in the boosted region $M_{t\bar{t}} \gg m_t$.

Besides QCD corrections, at high energies the electroweak (EW) corrections also become important \cite{Beenakker:1993yr, Bernreuther:2005is, Kuhn:2005it,  Bernreuther:2006vg, Kuhn:2006vh, Hollik:2007sw,  Bernreuther:2008md, Bernreuther:2010ny, Hollik:2011ps, Kuhn:2011ri, Manohar:2012rs, Bernreuther:2012sx, Kuhn:2013zoa, Campbell:2015vua, Pagani:2016caq, Czakon:2017wor, Czakon:2017lgo, Gutschow:2018tuk}. In \cite{Czakon:2017wor}, the complete next-to-leading order (NLO) corrections of QCD and EW origin are combined with the NNLO QCD results using the multiplicative approach (denoted as QCD$\times$EW in the following). The result shows that at high transverse momentum, the EW effects can significantly reduce the differential cross section, and need to be taken into account for an accurate modeling of the spectrum.

In this work, we perform a combination among four calculations for the differential cross sections in $t\bar{t}$ production:
\begin{enumerate}

\item The NNLO QCD calculation of \cite{Baernreuther:2012ws, Czakon:2012zr, Czakon:2012pz, Czakon:2013goa, Czakon:2014xsa, Czakon:2015owf, Czakon:2016dgf, Catani:2019iny, Catani:2019hip};

\item The soft gluon resummation of \cite{Ahrens:2010zv} at next-to-next-to-leading logarithmic (NNLL) accuracy;

\item The boosted soft gluon resummation of \cite{Ferroglia:2012ku, Pecjak:2016nee, Pecjak:2018lif} at NNLL$'$ accuracy;

\item The complete-NLO predictions of QCD and EW origin \cite{Czakon:2017wor, Gutschow:2018tuk, Frederix:2018nkq}.

\end{enumerate}
In this work, all of the currently-available perturbative contributions to these observables are combined. Therefore the results presented here are the state-of-the-art theoretical predictions within the SM. This is the first time that such a complicated combination of radiative corrections has appeared in the literature.

\section{Matching of the various corrections}
\label{sec:match}

In this section, we briefly introduce the four sets of corrections entering the combination, and then discuss the matching procedure employed to combine them. The matching procedure is necessary in order to remove the overlap among the various calculations.

To illustrate the idea, we discuss the matching procedure for the particular case of the invariant-mass distribution. The combination procedure for other distributions follows exactly the same pattern as the one for the invariant-mass distribution. One begins with the factorization formula
\begin{align}
\label{eq:xsec}
\frac{d\sigma(\tau)}{dM_{t\bar{t}}} = \frac{8\pi\beta_t}{3sM_{t\bar{t}}} \sum_{ij} \int d\Theta \int_\tau^1 \frac{dz}{z} \, \mathcal{L}_{ij}(\tau/z,\mu_f) \, C_{ij}(z,M_{t\bar{t}},m_t,\Theta,\mu_f) \, ,
\end{align}
where $M_{t\bar{t}}$ is the invariant mass of the $t\bar{t}$ pair; $\tau \equiv M_{t\bar{t}}^2/s$ and $z \equiv M_{t\bar{t}}^2/\hat{s}$ with $\sqrt{s}$ and $\sqrt{\hat{s}}$ being the hadronic and partonic center-of-mass energies respectively; $\beta_t = \sqrt{1-4m_t^2/M_{t\bar{t}}^2}$; $\mathcal{L}_{ij}(x,\mu_f)$ is the parton luminosity function with $\mu_f$ being the factorization scale; $C_{ij}$ is the partonic hard-scattering kernel where $\Theta$ was used to collectively denote additional kinematic variables. The sum in the above formula runs over initial-state partons $i,j=q,\bar{q},g$, and the prefactor is introduced by convention.

It is convenient to perform a Mellin transform of Eq.~\eqref{eq:xsec} with respect to $\tau$. After the transform, the differential cross section becomes
\begin{align}
\label{eq:xsec-mellin}
\frac{d\widetilde{\sigma}(N)}{dM_{t\bar{t}}} = \frac{8\pi\beta_t}{3sM_{t\bar{t}}} \sum_{ij} \int d\Theta \, \widetilde{\mathcal{L}}_{ij}(N,\mu_f) \, \widetilde{c}_{ij}(N,M_{t\bar{t}},m_t,\Theta,\mu_f) \, ,
\end{align}
where $N$ is the Mellin moment, and the symbols with a tilde denote the Mellin transform of the corresponding functions in Eq.~\eqref{eq:xsec}. In the following, we deal with the perturbative contributions to the hard-scattering kernel $\widetilde{c}_{ij}$ within the SM.

In fixed-order perturbation theory, $\widetilde{c}_{ij}$ can be expanded as a double series in the strong coupling constant $\alpha_s$ and the fine-structure constant $\alpha$. The NNLO QCD result contains the $\alpha_s^2$, $\alpha_s^3$ and $\alpha_s^4$ terms in the expansion; while the complete-NLO result includes the $\alpha_s^2$, $\alpha_s\alpha$, $\alpha^2$ terms at leading order (LO), and the $\alpha_s^3$, $\alpha_s^2\alpha$, $\alpha_s\alpha^2$, $\alpha^3$ terms at NLO. Such a fixed-order expansion is formally correct in generic phase-space regions. However, in certain kinematic limits, the fixed-order expansion breaks down due to the appearance of large logarithms at each order in perturbation theory. In such cases, especially in the case of pure QCD, the resummation of these logarithms is mandatory to avoid the bad convergence and/or the large scale dependence of the fixed-order results. 

One of the limits in which potentially large logarithms arise  is the threshold limit $z \to 1$ in momentum space, that corresponds to the $N \to \infty$ limit in Mellin space. In this limit the Mellin-space hard-scattering kernel $\widetilde{c}_{ij}$ develops large logarithms of the form $\alpha_s^nL^k$, where $L \sim \ln N$. The all-order resummation of these logarithms was studied in \cite{Kidonakis:1996aq, Kidonakis:1997gm, Ahrens:2010zv}. In \cite{Ahrens:2010zv}, the resummation was carried out at the next-to-next-to-leading logarithmic (NNLL) accuracy with the soft scale chosen in momentum space. The NNLL threshold resummation was re-evaluated in \cite{Pecjak:2016nee, Czakon:2018nun} with the soft scale chosen in Mellin space, in order to match the settings used in the boosted-soft resummation. Ignoring technical subtleties such as matrix-formed renormalization group (RG) evolution, the resummed hard-scattering kernel in Mellin space can be schematically written as
\begin{align}
\widetilde{c}_{ij} \sim \alpha_s^2 \, h_{ij}(\alpha_s) \, \exp \big[ g_{ij}(\alpha_s,\alpha_sL) \big] + \mathcal{O}(1/N) \, ,
\label{eq:resum}
\end{align}
for $ij=q\bar{q},\bar{q}q,gg$. All other partonic channels are power-suppressed in the $N \to \infty$ limit. The coefficient function $h_{ij}(\alpha_s)$ comes from the fixed-order calculation of the hard and soft functions \cite{Ahrens:2010zv}, while the exponent $g_{ij}(\alpha_s,\alpha_sL)$ comes from RG evolution. At NNLL accuracy, $h_{ij}(\alpha_s)$ needs to be evaluated up to NLO, namely, order $\alpha_s^1$. For the exponent $g_{ij}$, one counts $\alpha_sL \sim 1$ or $L \sim 1/\alpha_s$, and keeps the orders $\alpha_s^{-1}$, $\alpha_s^0$ and $\alpha_s^1$. By doing this, one resums on the exponent all terms of the form $\alpha_s^nL^m$, with $n-1 \leq m \leq n+1$. Upon expansion, this generates terms $\alpha_s^nL^k$ in fixed-order perturbation theory up to $k = 2n$. The various elements to achieve this level of logarithmic accuracy can be found in \cite{Ferroglia:2009ep, Ferroglia:2009ii, Ahrens:2010zv}. In this work, this result is denoted as NNLL$_{\text{m}}$, where the subscript `m' means ``massive'', in order to indicate that full dependence on the top-quark mass is retained.

In the threshold resummation framework discussed above, additional large logarithms of the form $\alpha_s^n\ln^l(m_t^2/M_{t\bar{t}}^2)$ ($l \le 2n$) might arise in the boosted limit $M_{t\bar{t}} \gg m_t$ or $\beta \to 1$. In this limit both the top and anti-top quarks are highly boosted in the $t\bar{t}$ rest frame. In \cite{Ferroglia:2012ku}, a framework was developed to simultaneously resum the two kinds of logarithms $\ln N$ and $\ln(m_t^2/M_{t\bar{t}}^2)$. The resummed result takes a very similar form to the one found Eq.~\eqref{eq:resum}, albeit with much more complicated functions $h_{ij}$ and $g_{ij}$. In addition,  $\mathcal{O}(m_t^2/M_{t\bar{t}}^2)$ power corrections are neglected in this boosted-soft resummation. With the ingredients evaluated in \cite{Ferroglia:2012ku, Ferroglia:2012uy, Ferroglia:2013zwa, Broggio:2014hoa}, the boosted-soft resummation was carried out at the NNLL$'$ accuracy in \cite{Pecjak:2016nee}, where the prime means that the coefficient function $h_{ij}$ has to be evaluated to one order higher, namely, to NNLO or order $\alpha_s^2$. The net effect of computing the $h_{ij}$ functions to one order higher is that the resummation captures the effect of one additional logarithm at each order in $\alpha_s$. In the following, we will denote this result as NNLL$'_{\text{b}}$, with `b' meaning ``boosted''.

\begin{table}[t!]
\centering
\begin{tabular}{|c|c|c|}
\hline
& \textbf{Included} & \textbf{Not included}
\\ \hline
NNLO QCD & $\alpha_s^n$ $(n = 2, 3, 4)$ & $\alpha_s^n$ $(n > 4)$; $\alpha_s^n\alpha^m$ $(m > 0)$
\\ \hline
NNLL$_{\text{m}}$ & $\alpha_s^n\ln^kN$ ($n \geq 2$) & $\alpha_s^n \mathcal{O}(1/N)$ ($n > 2$); $\alpha_s^n\alpha^m$ $(m > 0)$
\\ \hline
NNLL$'_{\text{b}}$ & $\alpha_s^n\ln^kN\ln^l(m_t^2/M_{t\bar{t}}^2)$ ($n \geq 2$) & $\alpha_s^n \mathcal{O}(1/N,m_t^2/M_{t\bar{t}}^2)$ ($n > 2$); $\alpha_s^n\alpha^m$ $(m > 0)$
\\ \hline
Complete NLO & $\alpha_s^n\alpha^m$ ($n+m = 2,3$) & $\alpha_s^n\alpha^m$ ($n + m > 3$)
\\ \hline
\end{tabular}
\caption{\label{tab:contributions}The contributions included in the four types of corrections entering the combination. See the text for a detailed explanation.}
\end{table}

To summarize, we collect in Table~\ref{tab:contributions} the contributions included (and not included) in the four types of corrections entering the final combination. One immediately sees that there are overlaps among them, particularly among the three QCD-based calculations. They need to be carefully removed in order to avoid the double-counting or even triple-counting of certain sets of corrections. This matching was achieved in \cite{Pecjak:2016nee, Czakon:2018nun} for the three purely QCD contributions. We first combine the NNLL$'_{\text{b}}$ and the NNLL$_{\text{m}}$ results to obtain an NNLL$'_{\text{b+m}}$ result. To do so, we need to remove the overlap between the NNLL$'_\text{b}$  and NNLL$_\text{m}$ results to all orders in $\alpha_s$. This can be done by exploiting the fact that the boosted-soft resummation formula is the small-mass limit of the soft-gluon resummation formula at any given order in $\alpha_s$. Therefore one finds 
\begin{equation}
\label{eq:res_comb}
d\sigma^{\text{NNLL}'_\text{b+m}} = d\sigma^{\text{NNLL}'_\text{b}} + \left( d\sigma^{\text{NNLL}_\text{m}} - \left. d\sigma^{\text{NNLL}_\text{m}} \right|_{m_t \to 0} \right) ,
\end{equation}
where the terms in the parenthesis account for contributions which are suppressed by $\alpha_s^nm_t^2/M_{t\bar{t}}^2$ for $n > 2$.

Subsequently, the matching with the NNLO QCD calculation proceeds by subtracting the NNLO expansion of the resummed formula
\begin{align}
\label{eq:fully-matchedNNLO}
d\sigma^{\text{NNLO+NNLL}'} = d\sigma^{\text{NNLL}'_\text{b+m}} + \bigg( d\sigma^{\text{NNLO}} - \left. d\sigma^{\text{NNLL}'_\text{b+m}} \right|_{\substack{\text{NNLO} \\ \mathrm{expansion}}} \bigg) \, ,
\end{align} 
where the terms in the parentheses account for contributions which are suppressed by $\alpha_s^n/N$ for $n=3,4$.

Finally, the complete-NLO contributions can be incorporated by first combining them with the NNLO QCD contributions in the multiplicative approach,\footnote{Orders $\alpha_s^n\alpha^m$ ($n+m \leq 3$) and $\alpha_s^4$ are summed and, bin-by-bin in any distributions, the order $\alpha_s^3\alpha$ contribution is approximated via rescaling the order $\alpha_s^2\alpha$ contribution by the NLO QCD $K$-factor. See Ref.~\cite{Czakon:2017wor} for more details.} arriving at the QCD$\times$EW result, and then matching against the resummation results as in Eq.~(\ref{eq:fully-matchedNNLO}). This leads to our final matching formula
\begin{align}
d\sigma^{\text{QCD} \times \text{EW} + \text{NNLL}'} = d\sigma^{\text{NNLL}'_\text{b+m}} + \bigg( d\sigma^{\text{QCD$\times$EW}} - \left. d\sigma^{\text{NNLL}'_\text{b+m}} \right|_{\substack{\text{NNLO} \\ \mathrm{expansion}}} \bigg) \, .
\label{eq:master}
\end{align}

\section{Numerical results}
\label{sec:num}

In this section, we present numerical results based on the matching formula Eq.~\eqref{eq:master}, and compare them with older predictions. For all the results we take the top-quark mass $m_t=\unit{172.5}{\GeV}$. Results for other top-quark masses can be obtained from the authors upon request. For purely QCD-based predictions we use the NNPDF3.1 NNLO PDF sets with $\alpha_s(m_Z)=0.118$ \cite{Ball:2017nwa}. When EW corrections are included, we use the NNPDF3.1 NNLO LUXQED PDF sets~\cite{Bertone:2017bme} with the same $\alpha_s(m_Z)$. There are a few of unphysical scales entering the fixed-order and resummed calculations. Their defaults choices in the results shown below are 
\begin{align}
\mu_r &= \mu_f = \begin{cases}
\frac{m_{T,t}}{2} \equiv \frac{1}{2}\sqrt{p_{T,t}^2+m_t^2} & \text{for } p_{T,t} \text{ distribution} \\
\frac{H_T}{4} \equiv \frac{1}{4} \left( \sqrt{p_{T,t}^2+m_t^2} + \sqrt{p_{T,\bar{t}}^2+m_t^2} \right) & \text{for all other distributions}
\end{cases} \, , \nonumber \\
\mu_h &= \frac{H_T}{2} \, , \quad \mu_s = \frac{H_T}{\bar{N}} \equiv \frac{H_T}{Ne^{\gamma_E}} \, , \nonumber
\\
\mu_{dh} &= m_t \, , \quad \mu_{ds} = \frac{m_t}{\bar{N}} \, .
\end{align}
For the meanings of these scales, we refer to \cite{Pecjak:2016nee, Czakon:2018nun}. Variations of the unphysical scales around the default values listed above are employed to estimate the impact of the higher order corrections that are not included in the calculations. Again, details on the scale variation procedure adopted can be found in  \cite{Pecjak:2016nee, Czakon:2018nun}.
For the settings on the EW parameters we refer to \cite{Czakon:2017wor}. The NLO EW calculation has been done within the latest public version of {\sc\small MadGraph5\_aMC@NLO} \cite{Frederix:2018nkq}.

\begin{figure}[t!]
\centering
\includegraphics[width=0.495\textwidth]{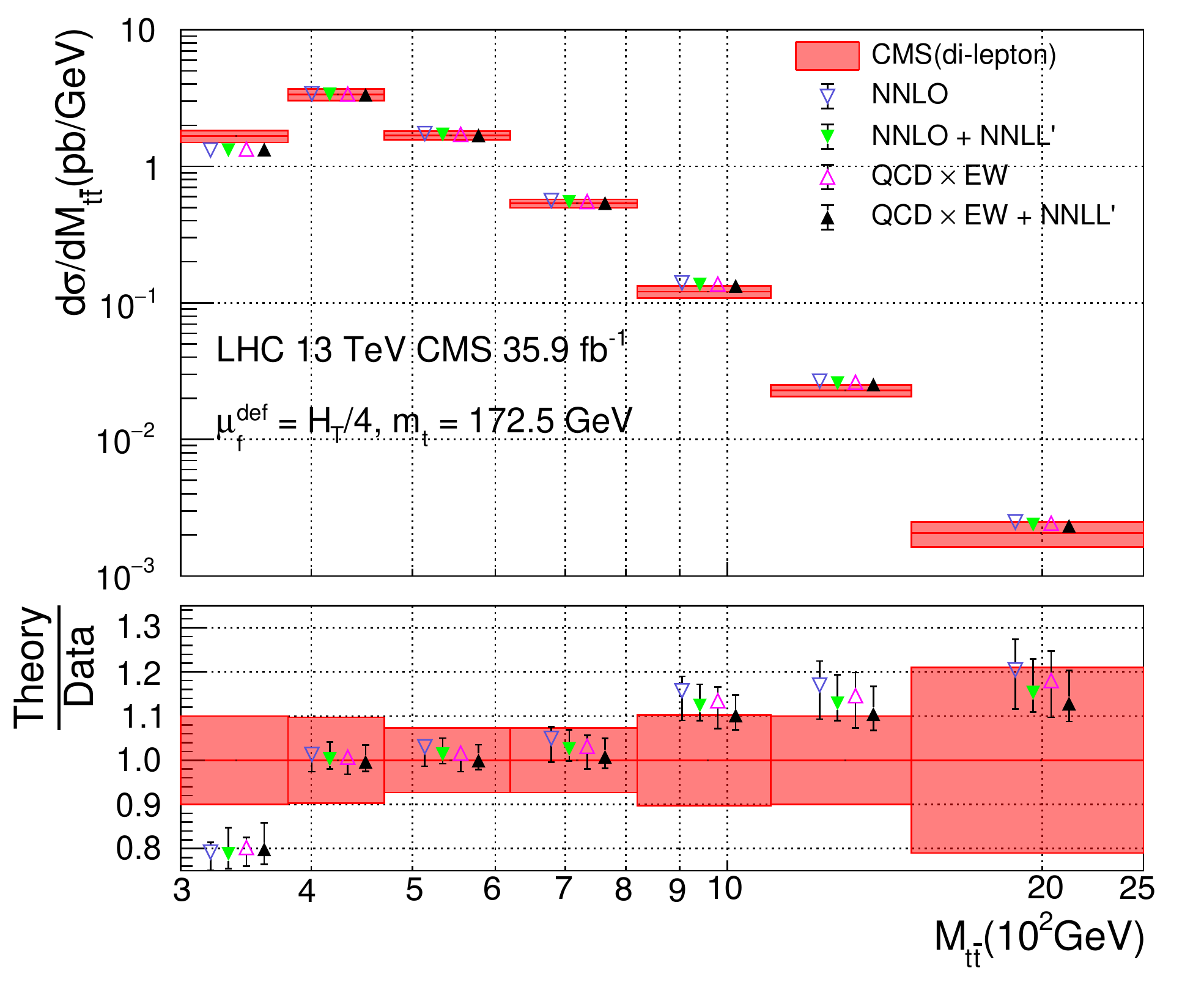}
\includegraphics[width=0.495\textwidth]{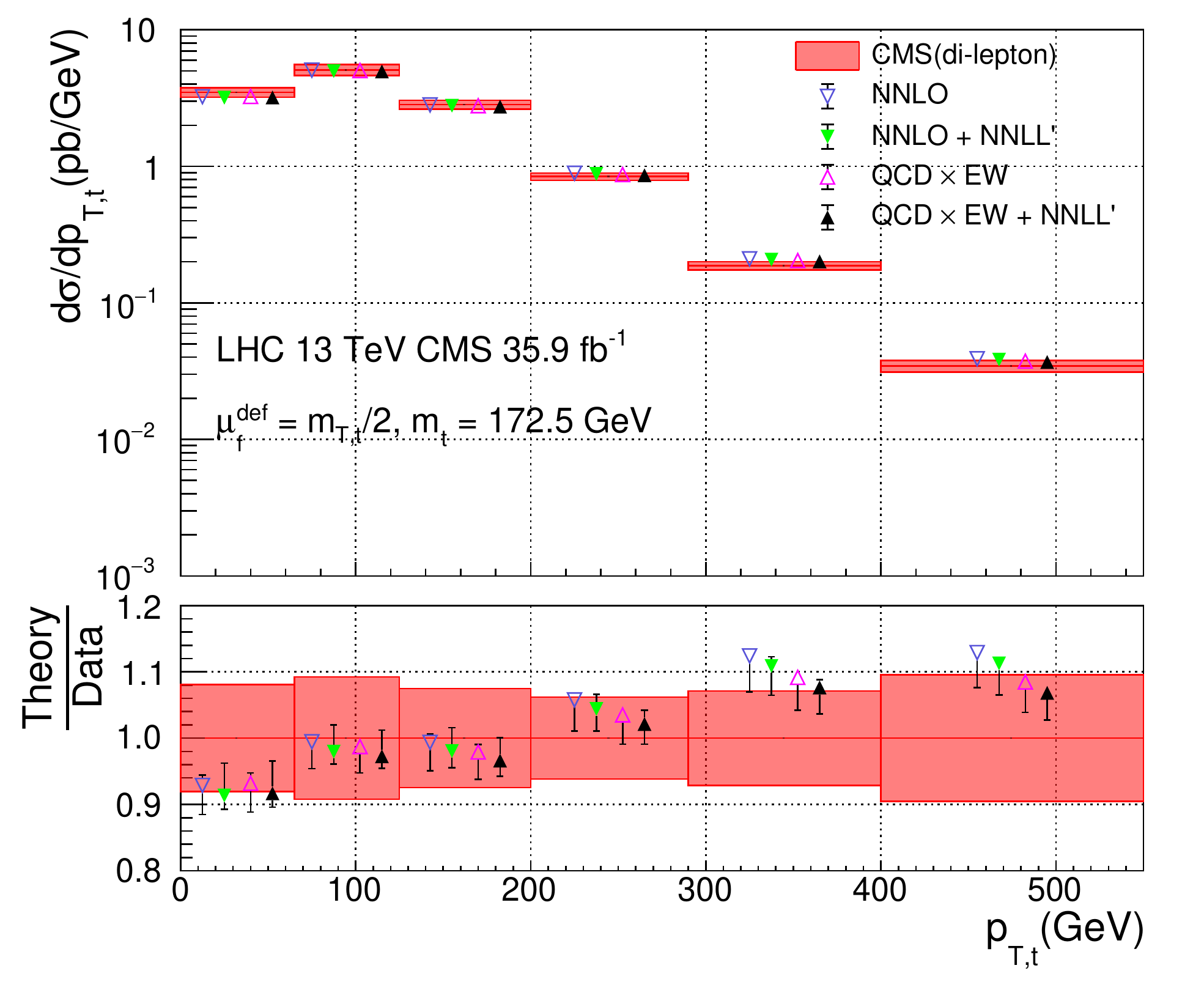}
\\
\includegraphics[width=0.495\textwidth]{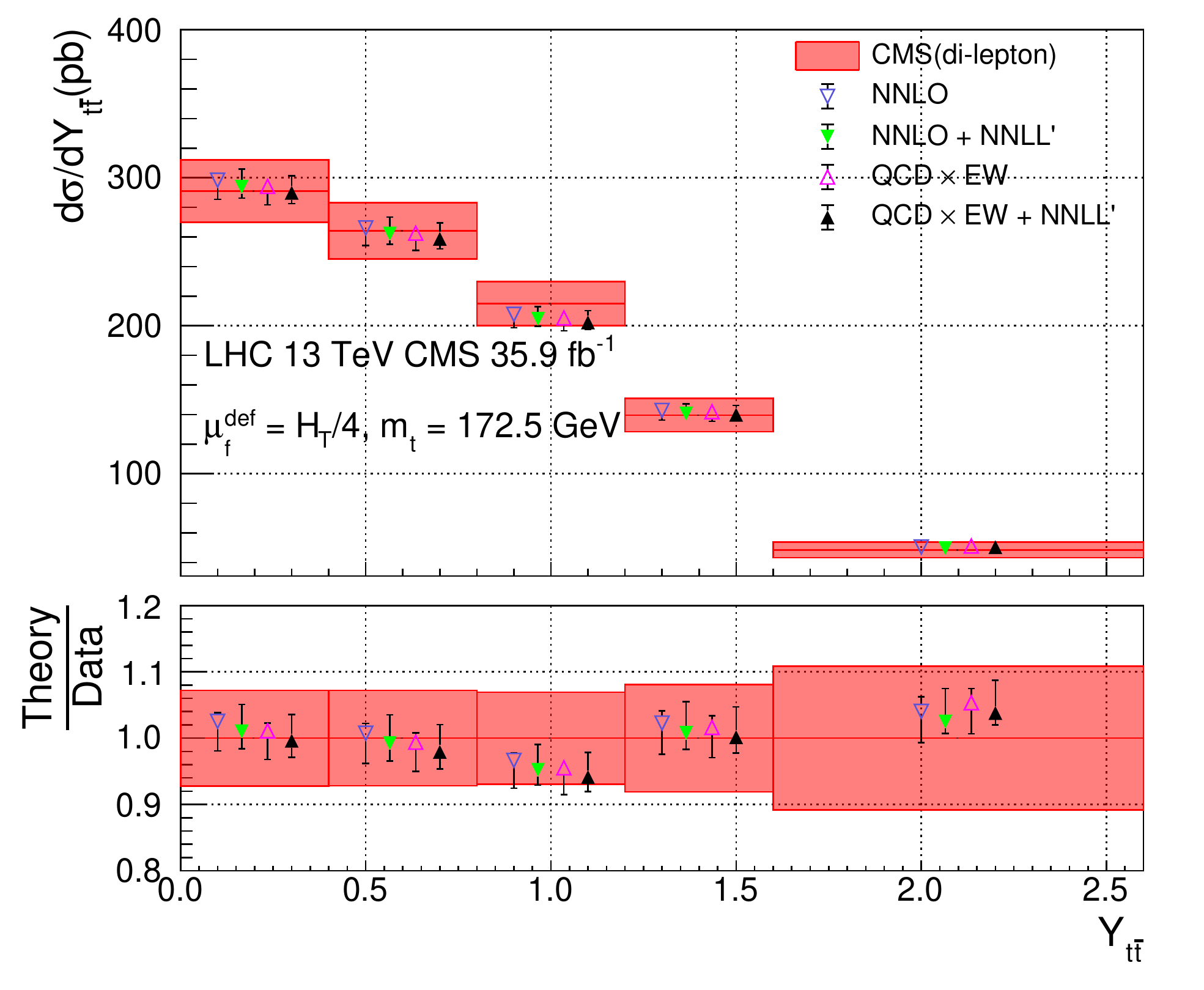}
\includegraphics[width=0.495\textwidth]{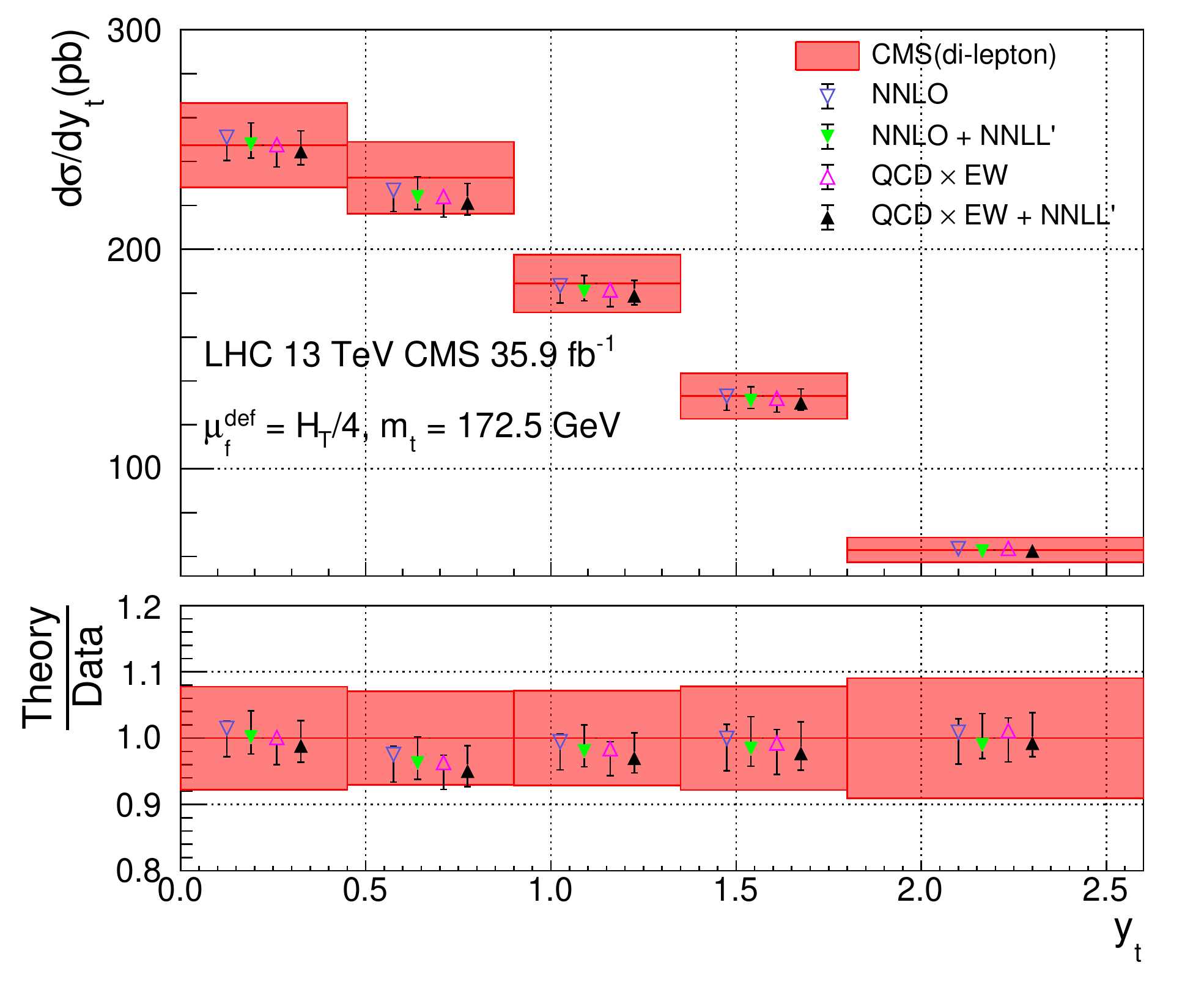}
\caption{\label{fig:4distswithEW}Theoretical predictions compared with the CMS data in the di-lepton channel \cite{Sirunyan:2018ucr}.}
\end{figure}

In Fig.~\ref{fig:4distswithEW}, we show predictions for  distributions differential with respect to
\begin{itemize}
\item[\emph{i)}] the $t\bar{t}$ invariant mass $M_{t\bar{t}}$, 
\item[\emph{ii)}] the transverse momentum $p_{T,t}$ of the top quark, 
\item[\emph{iii)}] the rapidity $Y_{t\bar{t}}$ of the $t\bar{t}$ pair, 
\item[\emph{iv)}] and  the rapidity $y_{t}$ of the top quark.
\end{itemize}
The four kinds of vertical bars correspond to the 4 kinds of theoretical predictions discussed in the last section: NNLO QCD, NNLO+NNLL$'$, QCD$\times$EW and QCD$\times$EW+NNLL$'$. The bands in red correspond to the CMS measurement in the di-lepton channel at the \unit{13}{\TeV} LHC using \unit{35.9}{\invfb} of data \cite{Sirunyan:2018ucr}.

By looking at the plots, one can conclude that the predictions are generically stable against inclusion of various sets of corrections. This signals that the convergence of the perturbative series and the estimate of the residual theoretical uncertainty affecting the predictions are well under-control. The effects of including QCD resummation and EW corrections are more evident in the large $M_{t\bar{t}}$ region and in the high $p_{T,t}$ tail (see inset in the first and second panel in Fig.~\ref{fig:4distswithEW}). In these cases, QCD resummation and EW corrections both tend to reduce the differential cross sections, which appear to be more compatible with experimental data than when those corrections are not included. In addition, the resummation effects enlarge the scale uncertainty in the first $M_{t\bar{t}}$ bin near the $2m_t$ threshold, where a small discrepancy is present between theoretical predictions and experimental measurement, thus slightly reducing the discrepancy. A recent study \cite{Ju:2019mqc} shows that a Coulomb resummation can significantly enhance the differential cross section in this region and can partly resolve the discrepancy.\footnote{Note that the differential cross section in the first bin is very sensitive to $m_t$. The discrepancy here decreases for smaller values of $m_t$.} Coulomb resummation can in principle be combined with the result in this work to achieve a good description of the $M_{t\bar{t}}$ spectrum in the whole phase space. For the rapidity distributions we can see that all the theoretical predictions provided lie almost completely within the uncertainty bands associated with the experimental measurements indicating excellent agreement for this observable. While the effect of the resummation on the uncertainty bands for these observables is minimal, we can still see that the effect of the higher order terms captured by the resummation is to slightly soften the rapidity spectrum compared to the corresponding fixed order predictions.

\section{Summary}
\label{sec:summary}

In this paper, we describe a combination among four calculations for the differential cross sections in $t\bar{t}$ production: the NNLO QCD calculations, the NNLL QCD threshold resummation, the NNLL$'$ QCD resummation for boosted top quarks, and the complete-NLO predictions of QCD and EW origin. This is the first time that such a complicated combination appears in the literature. The outcome represents the state-of-the-art prediction for $t\bar{t}$ differential distributions within the SM, which includes all sets of corrections available at the moment. Numerical results are presented for the invariant-mass distribution, the transverse-momentum distribution as well as rapidity distributions. We compare our predictions with the CMS measurements in the di-lepton channel at the \unit{13}{\TeV} LHC with an integrated luminosity of \unit{35.9}{\invfb}, and find overall good agreements.

\section*{Acknowledgements}

L.~L.~Yang and X.~Wang are supported in part by the National Natural Science Foundation of China under Grant No. 11975030, 11635001 and 11575004. D.~J.~S. is supported under the ERC grant ERC-STG2015-677323. The work of D.~P. and I.~T. is supported by the Alexander von Humboldt Foundation, in the framework of the Sofja Kovalevskaja Award Project ``Event Simulation for the Large Hadron Collider at High Precision''. D.~P. is also
supported by the Deutsche Forschungsgemeinschaft (DFG) under Germany’s
Excellence Strategy – EXC 2121``Quantum Univers'' – 390833306. I.~T. is also supported by the Swedish Research Council under contract number 2016-05996. The research of A.~M. and A.~P. has received funding from the European Research Council (ERC) under the European Union's Horizon 2020 research and innovation programme (grant agreement No 683211) as well as from UK STFC grants ST/L002760/1 and ST/K004883/1. The work of M.~C. was supported in part by a grant of the BMBF and by the Deutsche Forschungs-gemeinschaft under grant 396021762 - TRR 257.

\end{document}